\documentclass[journal=jpclcd,manuscript=letter,layout=twocolumn]{achemso}

\usepackage[utf8]{inputenc}
\usepackage[T1]{fontenc}
\usepackage{amsmath, amssymb}

\usepackage{graphicx}
\usepackage{float}       
\usepackage{tabularx}
\usepackage{multirow} 
\usepackage{makecell}

\usepackage{hyperref}
\usepackage{xr-hyper}    
\usepackage{cleveref}

\usepackage[dvipsnames]{xcolor}

\makeatletter
\newcommand*{\addFileDependency}[1]{%
  \typeout{(#1)}
  \@addtofilelist{#1}
  \IfFileExists{#1}{}{\typeout{No file #1.}}
}
\makeatother

\newcommand*{\myexternaldocument}[1]{%
    \externaldocument{#1}%
    \addFileDependency{#1.tex}%
    \addFileDependency{#1.aux}%
}

\myexternaldocument{si}

\newcommand{\bo}{B$_2$O$_3$}

\title{On the Boroxol Ring Fraction in Melt-Quenched B$_2$O$_3$ Glass}

\author{Debendra Meher}
\author{Nikhil V. S. Avula}
\author{Sundaram Balasubramanian}
\email{bala@jncasr.ac.in}
\affiliation{Chemistry and Physics of Materials Unit, Jawaharlal Nehru Centre for Advanced Scientific Research, Bangalore 560064, India}

\begin{tocentry}
\includegraphics[width=\linewidth]{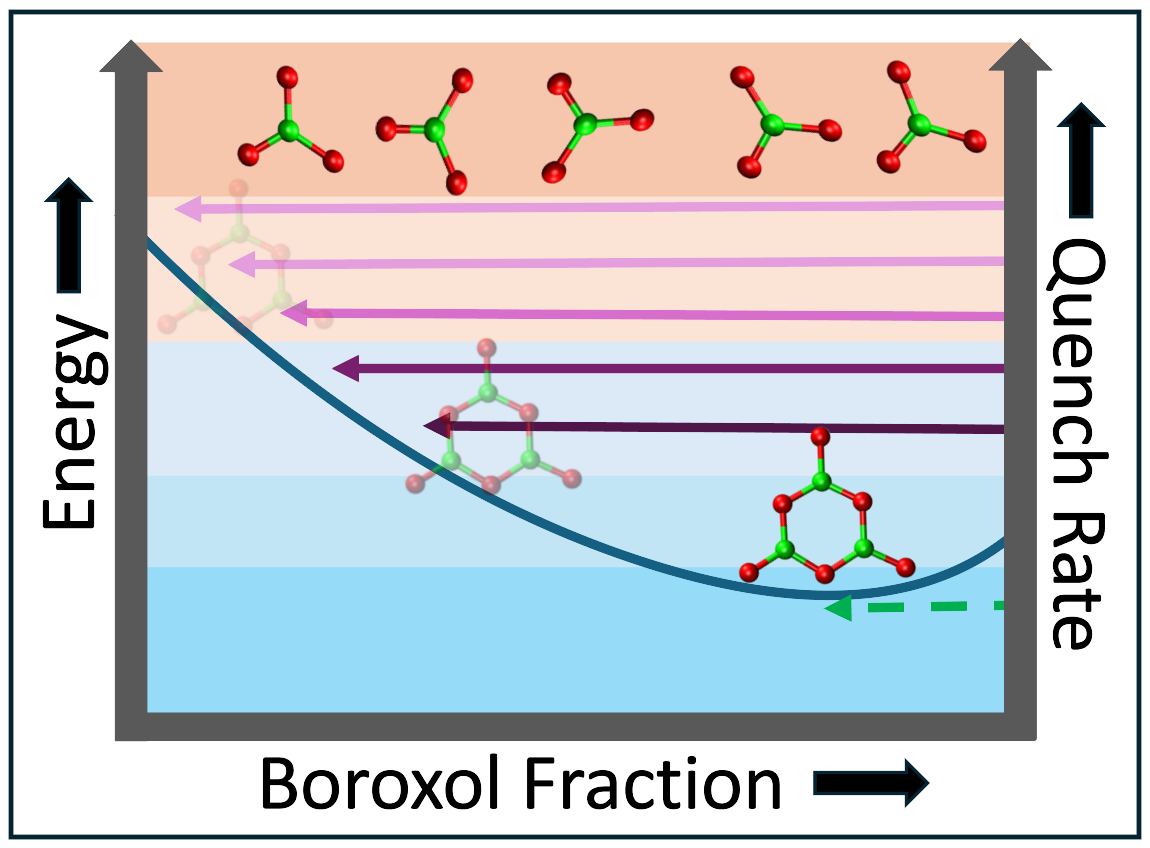}
\end{tocentry}

\begin{document}

\begin{abstract}
An atomistic structural model for melt-quenched \bo\ glass has eluded the simulation community so far. The difficulty lies in the abundance of the six-membered boroxol rings-- an intermediate-range order motif suggested  through Raman and NMR spectroscopy -- which is challenging to obtain in atomistic molecular dynamics simulations. Here, we report the development of a DFT-accurate machine-learned potential for \bo\ and employ quench rates as low as 10$^{9}$ K/s to obtain \bo\ glasses with more than 30\% of boron atoms in boroxol rings. Also, we show that the pressure, and consequently the boroxol fraction, in the deep potential molecular dynamics (DPMD) simulations critically depends on the range of the geometry descriptor used in the embedding neural network, and at least a 9\AA\ range is required. The boroxol ring fraction increases with decreasing quench rate. Finally, amorphous \bo\ configurations display a minimum in energy at a boroxol fraction of 75\%, intriguingly close to the experimental estimate in \bo\ glass.
\end{abstract}

\maketitle

\section{Introduction} 
The building block of boron trioxide (\bo) glass is the planar BO$_3$ unit\cite{MORNINGSTAR1936}. When three of these units come together to share an oxygen among each pair, a planar, six-membered ring, known as a boroxol ring\cite{ZACHARIASEN1032} is formed within its network. This hexagonal ring contains alternating boron and oxygen atoms (-B-O-B-O-) at its vertices. Raman spectroscopy and $^{[11]}$B NMR of \bo\ glass at ambient conditions (300 K, 1.834 g/cc) have shown that the fraction of boron atoms present in boroxol rings is 75\% \cite{KROGHMOE1969,Umari1374,expt_boroxol_1983} with the former displaying a sharp peak at 808 cm$^{-1}$ ~\cite{KELLER1953} assigned to the in-plane breathing mode of the ring. 

\begin{figure}
    \centering
    \includegraphics[width=1\linewidth]{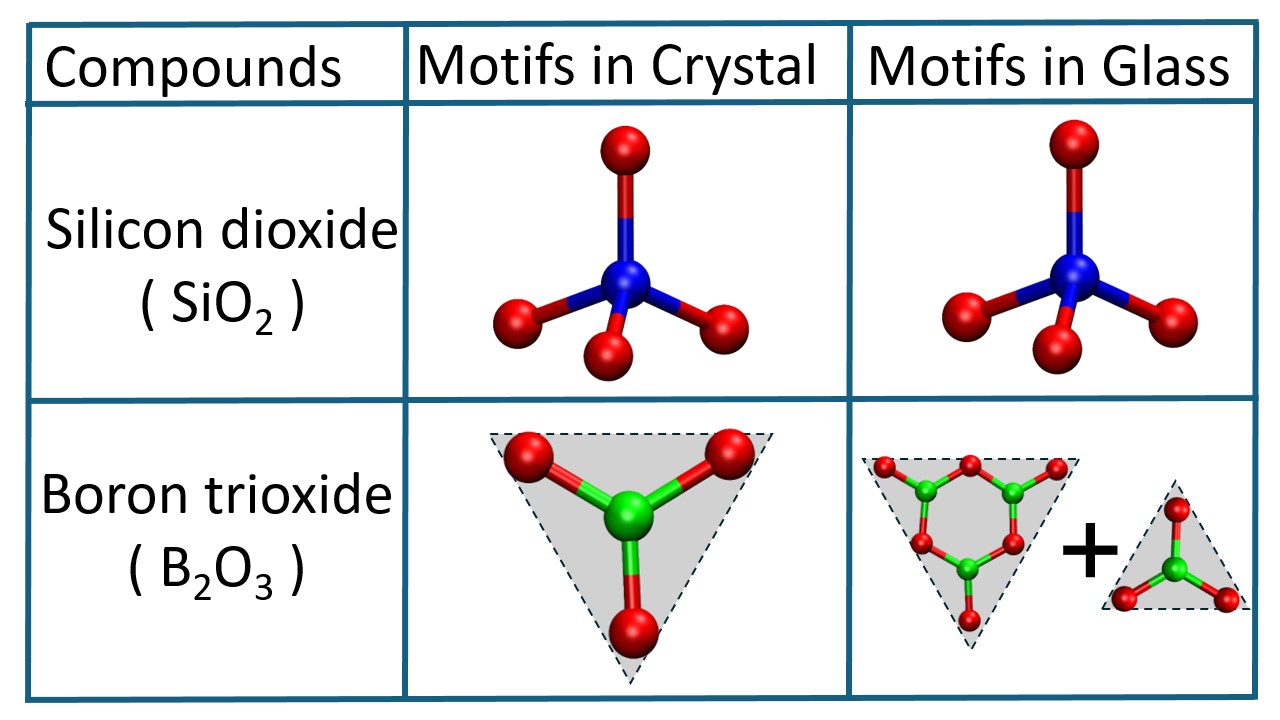}
    \caption{Comparison of boron trioxide with the common glass, silicon dioxide.}
    \label{fig:b2o3_sio2_comparison}
\end{figure}

In general, most glass systems and their corresponding most stable crystal structures at ambient conditions contain the same structural units. Consider silicon dioxide (SiO$_2$), for example; the smallest structural unit is the SiO$_4$ tetrahedron, which is present both in crystalline $\alpha$-quartz as well as in SiO$_2$ glass~\cite{sio2_1982}. This commonality makes it relatively easy to relate SiO$_2$ glass structures to the corresponding crystal structure.
In contrast, in \bo, the structural motifs of the glass and the crystal differ. Firstly, the experimentally known crystals of \bo\ \, \bo-I and \bo-II are crystallised at high pressures (1.5 GPa and 6.5 GPa respectively)~\cite{b2o3-I_1970,b2o3-II_1968}. No ambient pressure bulk crystal of \bo\ is known yet. 
Figure \ref{fig:b2o3_sio2_comparison} compares the structural motifs of SiO$_2$ and \bo\ in their crystalline and glassy forms. \bo-I\ contains only BO$_3$ structural units\cite{b2o3-I_1970,b2o3-II_1968}, while \bo-II contains tetrahedral BO$_4$ units~\cite{catlow1995}, with neither of them having hexagonal boroxol rings. In contrast, \bo\ glass consists of a mixture of boroxol (B$_3$O$_6$) and BO$_3$ units in the ratio of 3:1 ~\cite{Tian1997PRB,ferlat_prl_2008}. This fundamental aspect makes the atomistic modeling of \bo \ glass challenging. Its structure cannot just be 'extrapolated' from the corresponding crystal through, say, a disorder in the B-O-B angle over the network, à la SiO$_2$.
Further, while the density of SiO$_2$ glass is lower than that of $\alpha$-quartz by about 17\%~\cite{sio2_1982}, the corresponding value for \bo\ glass with respect to \bo-I\ is much larger, about 28\%~\cite{b2o3-I_1970}.
These differences in the density and the structural motifs between the glass and crystalline phases complicate the modeling of \bo\ glass using experimentally determined crystal structures (for instance).

Molecular dynamics (MD) simulations using a force field have contributed much to our understanding of 
structure and dynamics of glasses~\cite{Stillinger2001Polarization,anoop2018,Matthieu2014Polarization,Ferlat2019Polarization,MD_kiffer_2006,MD_kiffer_B2O3-0_2007}. See Supporting Information Section ~\ref{sec:FF_comparison} for a summary of results from literature using force 
field MD simulations to model \bo\ glass. Kob and coworkers~\cite{siddharth_kob_2020} employed an ingenious approach to derive an interatomic potential for modeling borate glasses with modifiers. It relies on both the results of ab initio MD simulations of the melt and the experimental properties of the glass. But, the potential is tied to a specific quench rate of 10$^{12}$K/s, making it unviable to study the formation of boroxol rings at low quench rates. 
Thus, the empirical nature of the interatomic interaction 
parameters lead to questions on their fidelity to the real-world sample, a lacuna that is aggravated 
by the challenges in model-independent and precise experimental determination of interatomic structure(s) for an amorphous system. 
On the other hand, ab initio MD (AIMD) simulations, based, say, on quantum density functional theory (DFT) do not contain empirical parameters and have been used to model glassy systems~\cite{Johnshe_2024,ferlat_prl_2008,Umari1374}. Yet, due to their computational complexity, they cannot generate trajectories beyond, say, 100 ps, making the sampling of atomic configurations in the viscous supercooled state impossible. 

Atomistic MD simulations based on Machine-Learned Potentials (MLP) have been able to address both the above 
concerns admirably. As they are based on a quantum ground truth (say, DFT), they are free of empirical 
parameters. Since they provide the total energy and atom forces for a configuration much faster than 
DFT, they are computationally tractable as well. Thus, MLP-based MD simulations are 
on an upswing across many areas of computational materials science and chemistry~\cite{JPCL2018Gabour,Morrow2024Angewchem, anjali2025,nikhil2023,MLP_takahiro_2024}. 

We had recently embarked on an effort to model glassy \bo~\cite{debendra} and, in particular, its structural motifs in and phase behavior at high pressures. It was based on an MLP (denoted as ML-26 in Ref.~\cite{debendra}), through which we observed the boroxol ring fraction at 300 K and 1.834 g/cc to be around ~15\% \cite{debendra}, a value much lower than the experimental estimate of 75\%\cite{KROGHMOE1969,Umari1374}. ML-26 was developed to understand the  structural evolution of \bo\ glass with pressure, and the following reasons could underlie its  underestimation of the boroxol fraction at ambient conditions:
\begin{itemize}
    \item The training set focused on configurations related to high-pressure glassy \bo\, where the boroxol fraction is low; thus, the ML model could not learn networks with a large fraction of boroxol rings.  
    \item The glass was obtained at a quenching rate of 1$\times$10$^{10}$ K/s, which is at least six orders of magnitude larger than the fastest quenching rate adopted in experiments. 
    \item The glass was obtained by quenching the melt under constant density conditions. Thus, the melt itself was modeled at a density of 1.834 g/cc, while the experimentally reported density of the \bo\ melt at 2000 K is 1.49 g/cc~\cite{Alderman_2025,NAPOLITANO1965}.
\end{itemize}

\section{Results}
Herein, we address these issues by enhancing the ML model (current version: ML-31~\cite{meher_2025_17559923}) through a substantial increase in the quantity of configurations rich in boroxol rings in the training dataset. Additionally, we applied various quenching protocols that mirror the experimentally 
reported temperature dependence of density. These efforts, along with a critical examination of the distance range in the embedding network of the Deep Potential (DP), allow us to obtain hitherto unattained high values of boroxol fraction in melt-quenched \bo\ glass.


\subsection{Validating the Machine Learning Potential}
\begin{figure}
    \centering
    \includegraphics[width=1\linewidth]{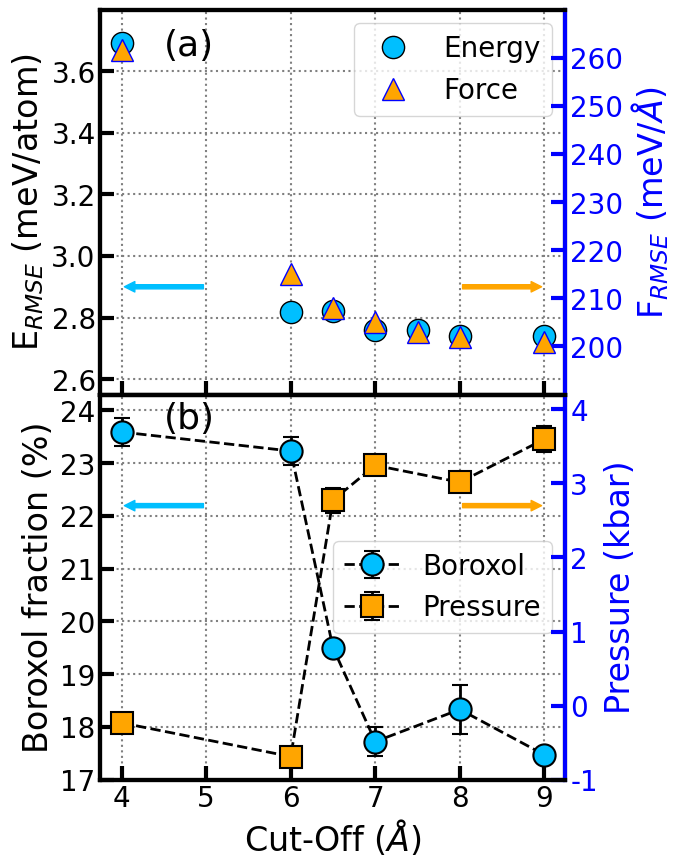}
    \caption{Performance of ML-31 MLP with different geometry descriptor cut-offs used in the embedding network. (a) RMSE of energy and the magnitude of force on the atom of the validation dataset versus cut-off. The size of the validation dataset is about 10\% of that of the training dataset. (b) Boroxol fraction and mean pressure of \bo\ melt simulated by ML-31 MLP-based DPMD at 2400 K vs. cut-off, modeled under constant-NVT conditions at a density of 1.49 g/cc, post an equilibration time of 10 ns. The results pertain to a system containing 1,700 atoms. Error bars represent the standard deviation calculated from four independent runs, using different initial configurations and velocity distributions. The pressure data for configurations generated through MD simulations with a model of a specific cut-off, but evaluated with a model of a different cut-off are presented in Figure~\ref{fig:virial_validaiton}.}
    \label{fig:boroxol_vs_mlp}
\end{figure}

Complete details of the steps involved in the development of ML-31 from ML-26 are provided in SI section~\ref{sec:mlp-refinment}. Here, we limit ourselves to some of its salient 
features.
Figure \ref{fig:boroxol_vs_mlp} (a) shows the relationship between the RMSE of energy and force for the validation dataset and the descriptor cut-off, a crucial step in the ML training process.  Clearly, a cut-off of 4 \AA\ is insufficient to capture the inter-atomic geometries in the configurations accurately. In contrast, cut-offs of 6 \AA\ (a value which is widely adopted for modeling molecular liquids~\cite{NNcutoff,anjali2025,nikhil2023}) and higher, yield similar RMSE values for energy and force, with small, but non-negligible improvements noted with increasing cut-off. The results begin to plateau around 8 to 9 \AA.

RMSE on energy and forces are standard checks on an ML model's fidelity; however, recently, various researchers have pointed out the need for additional tests~\cite{Klein2025MLP,Krishnan2025}. Interestingly, the pressure of the melt at 2400 K and 1.49 g/cc displays a strong dependence on 
the cut-off, in particular between 6-7 \AA. 
(Figure \ref{fig:boroxol_vs_mlp}(b)).  The temperature of 2400 K ensures sufficient sampling of configurations during the DPMD trajectory. The calculated pressure of the melt modelled by 
ML-31 models with different descriptor ranges converge only beyond a cut-off of 8 \AA. It has previously been suggested~\cite {Matthieu2014Polarization} that high pressures reduce the boroxol ring fraction in the system. The same is seen in Figure~\ref{fig:boroxol_vs_mlp}(b). The boroxol fraction and the pressure data display a crossover between 6-7 \AA\ distance cut-off. The exact structural motifs that cause this crossover need to be examined in the future.

Thus, a cut-off of even 6 \AA\ is insufficient to accurately capture the geometry of the atomic configuration. This suggests that significant interatomic structure beyond 6 \AA\ is present in glassy \bo\ and perhaps so in other inorganic glasses or melts, whose contribution to particularly the pressure is significant and cannot be captured in an effective manner with a shorter cut-off. Further, the energy of a configuration cannot be obtained by a structure descriptor whose range is 6 \AA; at least 9 \AA\ is required. Thus, when developing a Machine Learning Potential (MLP), care must be taken regarding the descriptor cut-off distance as well as other hyperparameters of the MLP. Further details can be found in Table~\ref{tab:MLP_hyper_parm}.


\subsection{Quenching Protocol} 
Apart from the descriptor cut-off range, the quench protocol adopted must also be examined for its 
impact on the glass structure. In MD simulations of inorganic glasses using empirical force fields, the standard procedure has been to simulate even the melt at the experimentally reported density of the glass and to quench the melt under constant-NVT conditions to obtain the glass at 300 K~\cite{NVT_quench1,NVT_quench2}. Even assuming that the interaction potential is accurate, the quenching of the melt under constant NPT conditions (say, at 1 atm pressure), will yield a 300 K glass whose density would depend on the quenching rate -- denser glasses at slower quenching rates. In fact, our earlier work~\cite{debendra} indirectly yielded the same conclusion on the quenching rate dependence of glass density. While constant-NPT simulations at each temperature could be adopted here, the calculated pressure of the system is tied to the adequate sampling of various network structures. As the structural relaxation times of the melt, as well as those of the supercooled liquid, are forbiddingly larger than the timescales accessible to atomistic MD simulations (see Figure~\ref{fig:boroxol_dviscosity_vs_density}), constant-NPT simulations with inadequate configurational sampling cannot be guaranteed to be an appropriate procedure. Thus, 
we adopt a slightly different strategy. 

In the case of \bo\ glass, the boroxol fraction will likely depend on the evolution of density with temperature. 
Indeed, Ferlat~\cite{Ferlat2015} states that using the 300 K glass density at high temperatures yields a rather high pressure, as expected.  Herein, we explore the possibility of the observed lower-than-experimental boroxol fraction at 300 K arising from modeling the high-temperature melt at a density higher than the experimental one. The experimental density vs. temperature curve~\ref{fig:expt_temp_vs_density} has been reported in literature~\cite{NAPOLITANO1965,Macedo1968}; it changes dramatically between 1200 K and 600 K. At temperatures below 600 K and above 1200 K, the variations in density are moderate. Thus, the constant-NVT quench at a density of 1.834 g/cc from high temperature to 300 K adopted so far can be reconsidered. One can still carry out a constant-NVT quench, but one that follows the experimentally reported curve as in Ref.~\ref{fig:expt_temp_vs_density} in a step-wise manner, while quenching at a constant rate. Toward this, firstly, we establish a random initial configuration containing 1700 atoms (340 \bo\ units) using Packmol~\cite{packmol,packmol2} with a box size corresponding to the experimental density of the melt at 2000 K (1.49 g/cc). After equilibrating it at this temperature, we cool the system over a specific duration by 50 K, and rescale the atom coordinates to correspond to the experimentally reported density at 1950 K. This process is repeated until 300 K and is elaborated through a schematic Figure~\ref {fig:nvt_expt_density}. This procedure of quenching will hereinafter be referred to as  NVT-$\boldsymbol{\rho}_{\text{EXP}}$ in this manuscript. 

\begin{figure}
    \centering
    \includegraphics[width=1\linewidth]{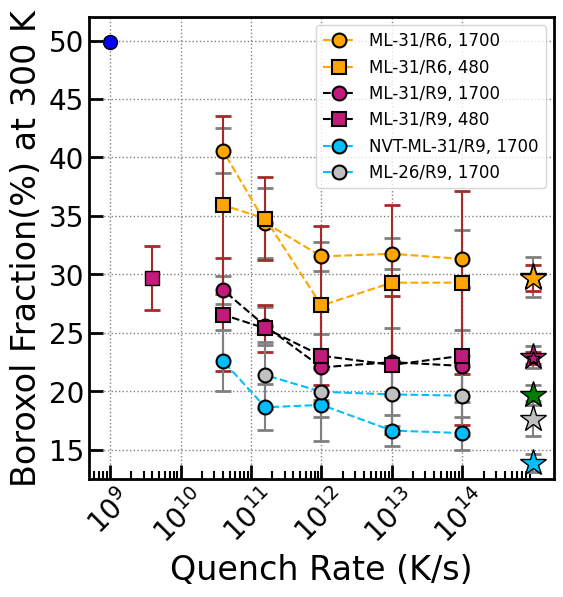}
    \caption{Dependence of boroxol ring fraction in the melt-quenched glass at 300 K on the quenching rate. Simulations were initiated from the melt at a temperature of 2000 K and a density of 1.49 g/cc, and followed the NVT-$\boldsymbol{\rho}_{\text{EXP}}$ procedure. Four independent simulations were conducted, each beginning with a different initial set of atom coordinates and velocities. Error bars represent the standard deviation calculated from these four runs. Star: Boroxol fraction in \bo\ melt at 2000 K, following a 10 ns equilibration period prior to quench. System sizes of either 1700 or 480 atoms were considered. The star in green colour is from a trajectory generated with the MACE~\cite{mace_arxiv} MLP trained on the ML-31 dataset. The single blue circle represents a single independent run with ML-31/R6, involving 1700 atoms and a quenching rate of $1\times10^{9}$ K/s, yielding a boroxol fraction of 50\% at 1200 K at the experimental density (1.537 g/cc). The data point for ML-31/R9 with 480 atoms at 4$\times10^9$ K/s is at 1400 K. Similarly, the data point for ML-26/R9 with 1700 atoms and quench rate $4\times10^{10}$ K/s is at a temperature of 1000 K. These values are unlikely to change by further cooling to 300 K. NVT-ML-31/R9, 1700 was obtained under constant-NVT quench, i.e., the melt at 2000 K and the liquid/glass during quench were maintained at the glass density of 1.834 g/cc. The blue circle took 90 days, while the rest of the data presented in this figure took an aggregate of 250 days of compute time on one A100 GPU card.}
    \label{fig:boroxol_diff_quench_rates}
\end{figure}

The above procedure adopted in DPMD simulations using the ML-26/R6 MLP yielded a boroxol fraction of just 19\% in the glass obtained from the melt at a constant quenching rate of 1.6$\times$$10^{11}$ K/s. Moreover, this method was not particularly helpful, as it requires spending a significant amount of simulation time at temperatures lower than 1400 K, at which conditions, the liquid is quite viscous. In fact, we note that below 1200 K, the boroxol fraction changes very little. Thus, we felt that it was prudent to expend the computational time for temperatures above 1200 K rather than below it.  Using ML-26/R6 at a quench rate 
of 1$\times$10$^{10}$ K/s from the 2000 K melt up to 1200 K, and an order of magnitude larger quench rate below 1200 K, resulted in a mean boroxol fraction of 30\% for the glass at 300 K. Thus, these results suggest that following the experimental density Vs. temperature data (in the absence of constant NPT DPMD simulations), an approach which is superficially superior to the constant NVT quench at the glass density, indeed enhances the boroxol fraction in the glass, as surmised. 

\begin{figure*}
    \centering
    \includegraphics[width=1\linewidth]{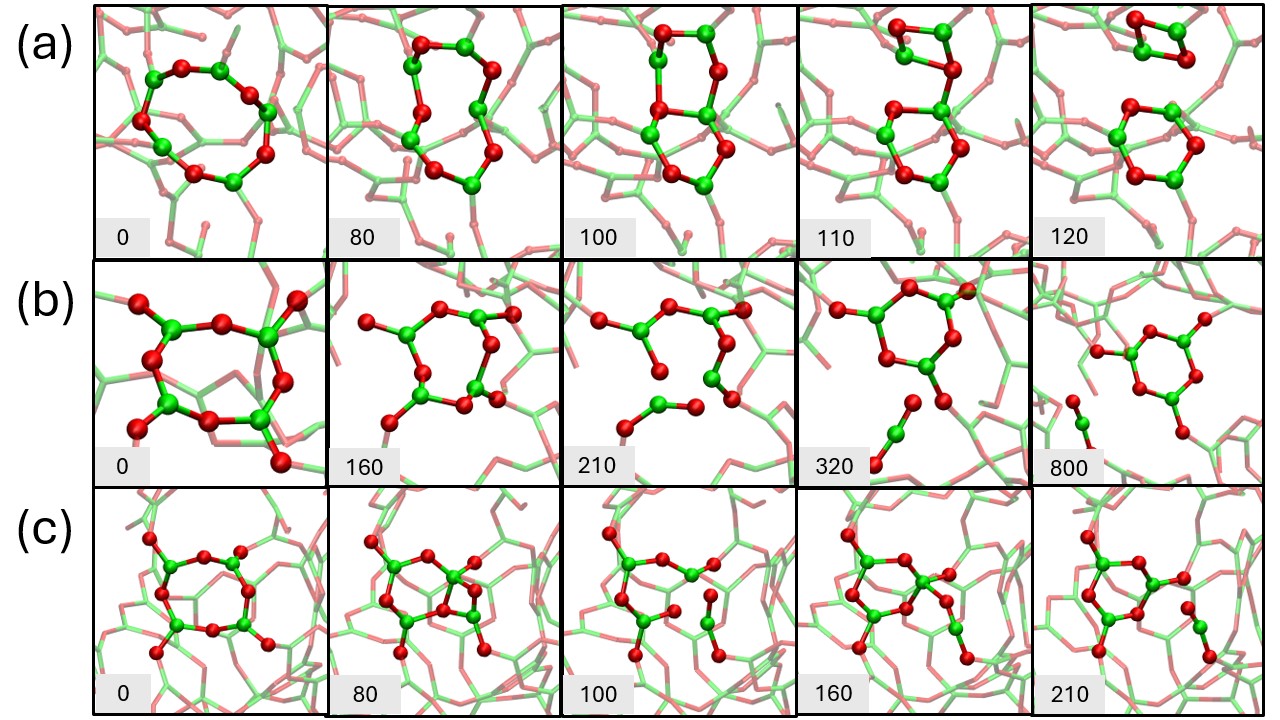}
    \caption{Arbitrarily chosen sequence of events during the formation of boroxol rings from different DPMD (ML-31/R9) trajectories containing 1700 atoms. Configurations in Panels (a) and (b) are from trajectories at 2000 K and 1.49 g/cc, while that in Panel (c) is from a trajectory at 1400 K and 1.515 g/cc. Green: Boron, Red: Oxygen. The numbers in the lower left corner of each configuration indicate the timestamp of the trajectory in femtoseconds. The zero in the timestamp is not the initial configuration, but rather a frame in a well-equilibrated trajectory.  Estimates of the duration between consecutive boroxol formation events (checked for every 20 ps) are 10 ps and 300 ps at 2000 K and 1400 K, respectively. Additional events are displayed in SI Figure~\ref{fig:boroxol_formed2}.}
    \label{fig:boroxol_mechanism}
\end{figure*}

Not only is the boroxol fraction at 300 K relevant, but its evolution from the melt with decreasing temperature also needs to be studied. 
Figure-\ref{fig:boroxol_ML31_run3} displays its growth as the melt is quenched at different rates using the ML-31/R6 version of the MLP. Decreasing quench rate increases the boroxol fraction in the glass, suggesting a large barrier for the nucleation of boroxol rings in the viscous \bo\ melt. Also shown in the figure is the behavior of the fraction with T, as obtained from experiments~\cite{boroxolvsT}. None of the growth curves obtained in the current set of DPMD simulations follows the experimentally reported one. The difference is likely due to the fact that the simulated samples increasingly move away from equilibrium as the temperature decreases. Another feature from the simulation data to note is the temperature-independent nature of the boroxol fraction below 1200 K.

Figure -\ref{fig:boroxol_diff_quench_rates} shows the boroxol fraction in \bo\ glass at 300 K obtained at different quenching rates starting from 2000 K. A striking feature of all the curves is they increase with decreasing quenching rate. Thus, it is not inconceivable that the boroxol fraction at much lower quenching rates, ones comparable to experimental rates, is 75\%. With 1700 atoms, the DPMD simulation carried out with ML-31/R9 MLP using the NVT-$\boldsymbol{\rho}_{\text{EXP}}$ procedure yields a larger boroxol fraction than the one 
done with the melt and its quench modelled at the glass density. This result reaffirms our 
earlier conclusion obtained with ML-26/R6 MLP.
Again, with 1700 atoms, DPMD simulations with 6 \AA\ cut-off yield a larger boroxol fraction than with 9 \AA. In fact, one simulation at a very low quench rate of 1x10$^9$ K/s yielded a glass structure with 50\% boroxol fraction. Decreasing the system size to 480 atoms, we could run four independent DPMD simulations with the ML-31/R9 model at a low quench rate of 4x10$^9$ K/s, which generated a \bo\ glass with
30\% boroxol ring fraction. However, the data for this system size is noisy, as even the formation (or destruction) of 
one boroxol ring led to a change in the boroxol fraction by about 2\%, due to the small system size. Keeping these in view, we believe that the results from the lowest quench rate obtained for ML-31/R9 model with 1700 atoms, of 30\% boroxol fraction, are the most reliable and robust ones.

\subsection{Boroxol Formation Mechanism}
In Figure~\ref{fig:boroxol_mechanism}, we exhibit three arbitrarily chosen illustrations of the formation of boroxol rings through the reorganization of the \bo\ network. Larger rings break into smaller ones through a short-lived BO$_4$ species (four-coordinated boron atom), which appears to be the dominant route to the formation of the six-membered rings, based on visual inspection. Although the interval between the formation of consecutive boroxol rings is large, the reorganization of the network leading to the boroxol ring formation happens relatively fast, within a picosecond or so. 

The reorganization of a generic n-membered ring and one that leads to the formation of a six-membered boroxol ring proceeds as follows: A large ring gets pinched, leading to the momentary formation of a four-coordinated boron atom as well as a three-coordinated oxygen atom. These vertices act as 'budding centers' which then split up, resulting in the six-membered boroxol ring within the \bo\ network. As discussed in SI Section~\ref{sec:structral_analysis}, ML-31 possesses the capability of capturing the geometry and energetics of 4-coordinated boron atoms well, as amorphous configurations at high densities were also included in its training set. This attribute, in hindsight, is crucial for describing the formation of boroxol rings in the network during the slow cooling of the melt. An examination of the ring statistics 
shown in Figure~\ref{fig:network_dist} reveals that an increased formation of boroxol rings in the melt-quenched glass is accompanied by a decrease in the proportion of 10- and 14-membered rings; the analysis also suggests a marginal, yet systematic increase in the proportion of 30-membered rings.

The classic signature of the presence of boroxol rings in \bo\ glass is the intense peak at 808 cm$^{-1}$ observed in Raman spectroscopy~\cite{KELLER1953,raman2}, which is assigned to the breathing mode of the six-membered ring. While we could calculate the Raman spectra using density functional perturbation theory (DFPT) for the in-silico generated crystalline configurations (see Section ~\ref{sec:raman_vdos} of SI), the same method could not be adopted for the glassy configurations obtained from melt-quenching herein, as the number of atoms was much larger (either 480 or 1700 atoms) than those considered for the crystalline states, resulting in insurmountable computational challenges. Hence, instead of the Raman spectrum, we have calculated the power spectrum, also known as the vibrational density of states (VDOS), as the cosine transform of the velocity auto-correlation function (VACF). These were calculated from fresh trajectories at 300 K temperature and 1.834 g/cc density, generated for a duration of 20 ps, while saving the velocities of all the atoms every 1 fs. 

\begin{figure}
    \centering
    \includegraphics[width=1\linewidth]{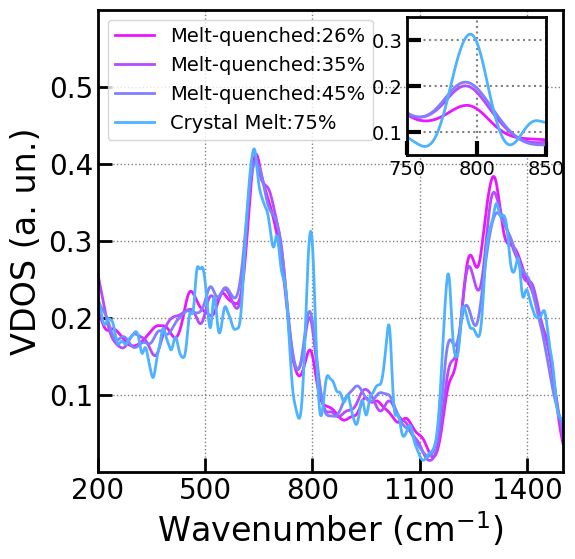}
    \caption{Vibrational density of states of amorphous \bo\ configurations containing different boroxol ring fractions. Inset displays the region around 800 cm$^{-1}$, whereas an intense Raman peak is observed experimentally at 808 cm$^{-1}$. The VDOS data has been smoothened using a Gaussian function with a half-width of 10 cm$^{-1}$.}
    \label{fig:vdos_with_diff_boroxol}
\end{figure}

The same is displayed in Figure~\ref{fig:vdos_with_diff_boroxol} for four different samples containing varied amounts of boroxol rings; three of these were obtained by melt-quenching, while a fourth was obtained by amorphization of the in-silico generated crystal, B$_2$O$_3$-T3-b. 
The VDOS shows a sharp peak at 790 cm$^{-1}$, which fares well against the experimental peak position of 808 cm$^{-1}$. Importantly, the peak intensity systematically increases with an increase in boroxol content, affirming its origin. 


\subsection{Boroxol Stability}
The DPMD simulation results on boroxol fraction discussed thus far can be summarized as follows. (i) it increases with decreasing temperature, (ii) with decreasing quenching rate, the fraction in the glass increases, and values ranging between 30\% and 40\% have been obtained in the current work. These two observations strongly suggest that amorphous structures containing an increasing boroxol fraction are more stable than ones with lower boroxol content. 
Given that the experimental glass is reported to contain a boroxol ring fraction of 75\%~\cite{KROGHMOE1969,Umari1374,expt_boroxol_1983}, it is pertinent to study the total energy of amorphous structures as a function of this fraction.  

Toward this aim, we generated 500 amorphous \bo\ configurations
whose boroxol fraction spans from 10\% to 100\%; the procedure followed to arrive at these structures is depicted in the schematic Figure~\ref{fig:boroxol_frm_collections}. In short, these configurations are derived from crystal structures generated in silico by Ferlat et al. ~\cite{Ferlat2012}, which contain different boroxol fractions following a process of amorphization. All these 500 configurations were maintained at a density of 1.834 g/cc, close to the density of vitreous \bo\ at room temperature reported experimentally (~1.83 g/cc)~\cite{NAPOLITANO1965,Alderman_2025}. X-ray and neutron structure factors (Figures \ref{fig:xray_structure}-\ref{fig:neutron_structure}) computed for these configurations show good agreement with experimental data, thereby confirming their amorphous nature.

\begin{figure}
    \centering
    \includegraphics[width=0.75\linewidth]{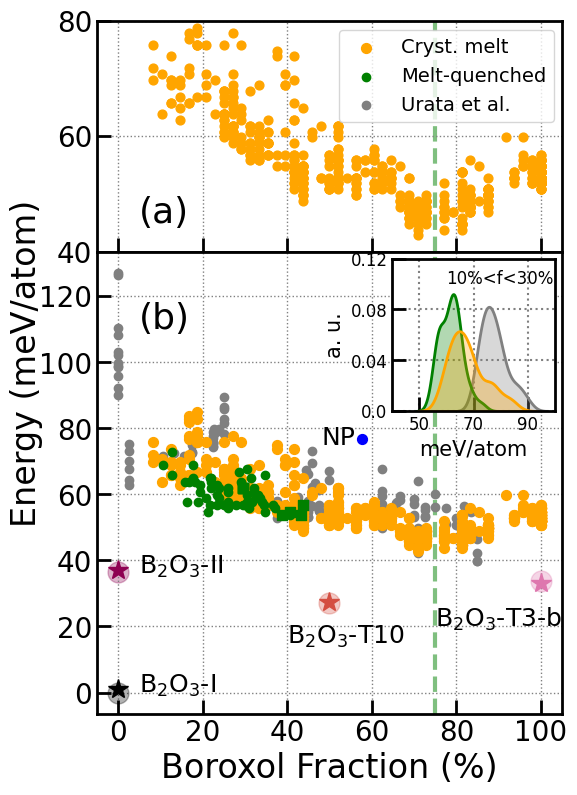}
    \caption{(a) Energy per atom of amorphous \bo\ configurations obtained after geometry optimization of the structure. Orange: 500 data points, each containing 360 atoms at 1.834 g/cc. These configurations were derived from the in-silico generated T3-b crystal\cite{Ferlat2012} that contains 100\% boroxol rings through the procedure presented in Figure~\ref {fig:boroxol_frm_collections}. Green: Sixty data points each containing 1700 atoms obtained by the melt-quenched method starting from 2000 K, while following the experimental density-temperature curve. Green circles: 1.834 g/cc; Green squares: 1.57 g/cc.  Blue: Single configuration with 1700 atoms at 1.834 g/cc obtained after applying negative pressure to the 2000 K melt, followed by quenching; further details can be found in Sec.-\ref{sec:neg-press}. Grey: 105 data points collected from Urata et. al.~\cite{URATA2025102474}, wherein the system sizes are 180, 300, or 3000 atoms, all at a density of 1.843 g/cc. Vertical dashed line marks the boroxol ring fraction of 75\%, where an energy minimum is seen. (b) Same as (a) but with additional data points from crystalline configurations, after geometry optimization. Crystal density in g/cc in parantheses: \bo\ -I (2.56), \bo\ -II (3.11) g/cc,  \bo\ -T10 (1.49), \bo\ -T3-b (0.93). The former two are experimentally reported ones, while the latter two are in-silico generated~\cite{Ferlat2012}. Transparent circles behind the star marks are energies calculated using DFT. Inset: Distribution of atom energies for the three sets of amorphous 
    configurations whose boroxol ring fraction, f, lie between 10-30\%.}
    \label{fig:boroxol_vs_nrg}
\end{figure}

As the boroxol fraction in a configuration increases, the energy of the system decreases. This observation aligns with previous first-principles molecular dynamics studies, which reported a monotonic decrease in energy with increasing boroxol fraction. Specifically, it was found that boroxol-rich structures are more stable than boroxol-poor ones at the glass density, with an approximate energy gain of $6.6 \pm 1~\mathrm{kcal/mol.B_2O_3}$ ~\cite{ferlat_prl_2008}. The corresponding value determined by us (Figure~\ref{fig:boroxol_vs_nrg}) is around $3 \mathrm{kcal/mol.B_2O_3}$. At the glass density of 1.834 g/cc, a low value relative to the crystalline form, lower 
energy amorphous structures can be attained only through the formation of boroxol rings. 
Panel (a) of Figure~\ref{fig:boroxol_vs_nrg} displays their energies obtained after structure optimization. The MACE model was used as a surrogate for DFT to obtain the energy, as its accuracy is excellent (see its parity plot in Figure~\ref {fig:mace-parity}) and can be evaluated at a significantly reduced computational cost. Also shown in the panel are the energies of sixty additional configurations (in green color) obtained by the melt-quench method. While the data in orange refer to configurations at a density of 1.834 g/cc, most of the ones in green are at the same density, while a few are at a lower density of 1.57 g/cc. These configurations, too, were 
geometry optimized prior to the calculation of their energies. Urata and Lodesani recently made an effort to estimate the boroxol ring fraction in \bo\ glass, and 
in the process examined amorphous configurations with varying boroxol fractions~\cite{URATA2025102474}. The energies of these configurations, too, after geometry optimization, are presented in Figure~\ref{fig:boroxol_vs_nrg}.
Consistent with the increase in boroxol ring fraction with decreasing temperature reported in our melt-quenching simulations, we note that the energies of these amorphous configurations also display a decreasing trend with increasing boroxol fraction.

Interestingly, Figure~\ref{fig:boroxol_vs_nrg}(a)  displays a minimum in energy at around 70-75\% boroxol, suggesting that configurations in this region represent the most stable amorphous structures. This finding happens to coincide with the prevailing consensus regarding the boroxol ring fraction in vitreous \bo\, which is widely reported to be around 75-80\% based on various techniques, including Raman and NMR spectroscopy, and neutron diffraction\cite{Alderman_2025}. Our observation that increasing the boroxol fraction beyond this optimum range (i.e., above 75\%) leads to higher energy configurations further supports the existence of an energetically preferred structural arrangement. Boroxol 
formation, while lowering the total energy, also leads to more porous structures, as seen from the in-silico generated crystal structures of Ferlat~\cite{Ferlat2012}, all of whose densities are lower than that of the \bo\ glass. Given that all the configurations denoted by orange circles in Figure~\ref{fig:boroxol_vs_nrg}(a) are at the same density of 1.834 g/cc, it is likely that the boroxol-rich amorphous configurations are strained so as to meet the bulk density constraint. This argument explains the presence of a minimum energy.

This successful differentiation by our MLP is significant, as empirical force field based molecular dynamics simulations often struggled to reproduce the boroxol fractions unless specific features like oxygen polarisation~\cite{Stillinger2001Polarization,Matthieu2014Polarization,Ferlat2019Polarization} or exceptionally slow quench rates or boroxol-rich initial configurations were incorporated. The ability of the MACE MLP to resolve these subtle energetic distinctions underscores its accuracy in modeling the complex intermolecular forces and structural preferences in vitreous \bo.

 Included in panel (b) of Figure~\ref{fig:boroxol_vs_nrg} are additional data --  the energies of the \bo-I and \bo-II crystals and those of the in-silico generated, low-density crystals, \bo-T10, 
and \bo-T3-b introduced by Ferlat et al~\cite{Ferlat2012}. \bo-I has a significantly higher density (2.56 g/cc) than the ambient pressure glass, as well as those of the computer-generated crystals with high boroxol fractions (whose density ranges from 0.93 to 1.49 g/cc).  As expected, the energy of the \bo-I\ crystal is considerably lower than that of any of the glass system configurations, including those with a high boroxol fraction. The relative stability of the \bo-I crystal, despite having zero boroxol rings, has two contributions: crystallinity and higher density, of which the latter is likely to be more dominant. The inset in panel (b) displays the distribution of atom energies for configurations whose boroxol ring fraction, $f$ lie between 10-30\%. Configurations obtained through melt-quenching procedure show the greatest stability, while those from amorphization of the in-silico crystal structures show a wide range of energies. This indicates that the boroxol fraction alone does not govern the relative stability of the glass, underscoring the influence of the preparation protocol. The data suggests that ultrastable \bo\ glass through molecular simulations is not unattainable. 

These results demonstrate that the ML-31 model is capable of distinguishing the energy differences between amorphous configurations, all at the same density, but with different boroxol fractions. Further examination of low-energy amorphous and crystalline configurations across the boroxol-density domain through generative AI methods could potentially shed light on the \bo\ crystallization anomaly, an effort that we plan to pursue in the future.


\section{Discussion}
Our earlier work focused on the study of high-pressure glassy \bo\ (boron trioxide) and specifically the transformation of trigonal planar BO$_3$ units to the tetrahedral BO$_4$ units with increasing pressure~\cite{debendra}. The training set for the MLP developed for this purpose (ML-26) contained amorphous configurations with densities ranging from approximately 0.9 g/cc to 6.2 g/cc. It enabled us to capture the properties of high-pressure glass and reproduce the glass structure at ambient conditions with very good accuracy compared to experimental results. 
As has been documented earlier by several researchers~\cite{BIONDUCCI1994RMC,Swenson1997RMC,Verhoef01111991,VERHOEF1992267,Enciso1996MD,Bermejo1996,Senent1996}, an apparent good agreement of computations with experimentally determined X-ray and neutron scattering structure factors, while necessary, is not sufficient to validate the boroxol ring fraction observed in the simulation. 

The current work is devoted to the examination of the fraction of boron atoms present within hexagonal boroxol (B$_3$O$_3$) rings in melt-quenched \bo\ glass at ambient conditions. Estimates from both Raman and NMR measurements indicate that this fraction is 75\%. However, MD simulations employing the melt quench protocol have not so far been able to reproduce this value, possibly arising from (a) inadequacies in the interatomic potential, (b) at least six orders of magnitude higher quench rates adopted in simulations, (c) the temperature dependence of density during quenching being different in computations from experiment. 

Herein, by employing accurate and systematically tested machine learned potentials, we have addressed the factor (a) above. Furthermore, by following the density-temperature curve of experiments in a stepwise manner, we have also considerably addressed factor (c). A high quench rate (which, in turn, manifests as insufficient sampling) is germane to atomistic MD simulations, and we have addressed it to the best extent possible, reaching quench rates hitherto unreported for inorganic glasses modelled at DFT accuracy, of 4x10$^{9}$K/s. In this manner, we have been able to demonstrate that (i) the boroxol fraction increases with decreasing quench rate and (ii) the fraction at the lowest quench rate for the ML-31/R9 model is 30\%. Our preliminary variable quench rate simulations suggest the possibility of obtaining a glass configuration with a higher boroxol fraction in a computationally efficient manner. See SI Section~\ref{sec:var_quench_rate} for details.

The process of MLP development to model ambient pressure \bo\ glass involved discarding several high-pressure configurations that were present in the training set for ML-26, the version of the MLP reported by us earlier~\cite{debendra}. Further, several amorphous configurations derived from computer-generated crystalline structures with varying amounts of boroxol fraction were included to constitute the training set for ML-31, as deployed in the present work. The MACE model with a 6 \AA\ geometry descriptor range reproduced DFT energies excellently as it intrinsically includes many-body correlations. On the other hand, the DP model with the same range displayed acceptable, yet larger RMSE values in both energies and atom forces. Marginally lower RMSE values with the DP model were attained at larger descriptor ranges, such as 9 \AA. A consequence of the use of a short-range (6 \AA) in the DP model is a significantly reduced pressure value compared to values obtained at higher cut-off ranges. which is partially offset by the longer-range value of 9 \AA. Interestingly, the reduction in pressure leads to an overestimation of the boroxol fraction in the configurations. Thus, an accurate representation of intermediate range order in \bo\ glass vitally depends on the range of the geometry descriptor used in the DP framework. The boroxol ring is smaller in size than the minimum descriptor range required to have accurate atom forces and virial. Thus, 
having a good number of configurations with varied amounts of boroxol fraction, while necessary to build a decent MLP, is not sufficient. The descriptor range, particularly in MLP frameworks such as the DP, is a vital component to maintain fidelity to the underlying DFT potential energy surface.

Although the experimentally reported 75\% boroxol fraction in \bo\ glass could not be achieved in any of the current set of MD simulations, the work has been able to narrow down the reason for the same to the high viscosity of the melt even at temperatures as high as 1600 K (see Figure \ref{fig:boroxol_dviscosity_vs_density}). Even the slowest quench rates employed here (4x10$^{9}$ K/s) are not sufficient to equilibrate the sample at these conditions. While methods to enhance sampling to produce ultrastable glasses~\cite{Berthier2024SMC,Ediger2019Vapourdiposition,Raegen2020}, such as the swap Monte Carlo, physical vapor deposition etc. exist, they have been applied mostly to Lennard-Jones glasses, and not for a network glass former such as \bo. Efforts to pursue such methods are in progress in our 
laboratory. Denoising-based diffusion models, such as the AMDEN reported recently~\cite{finkler2025amden}, too could be attempted to 
generate amorphous \bo\ configurations which meet all the known experimental properties, including the boroxol fraction. The distribution of DFT energies of such configurations and their comparison to those of the melt-quenched glass configurations reported here could be interesting.  The recent work of Zio et al~\cite{zio_science_advances_2025} on the growth of a 
two-dimensional boroxol-rich crystal on a substrate by atom deposition is also relevant to the discussion on 
boroxol rings being an essential requirement for the stability of bulk \bo\ glass.  Very recently, 
a crystal structure of B$_2$O$_3$ sheets intercalated with methanesulfonic acid has been reported~\cite{Logemann2025}. The sheets are constituted by 
boroxol rings and display the characteristic 808 cm$^{-1}$ Raman band.
Enhanced sampling using a machine-learned collective variable, such as that used recently to study the phase transition in elemental sulfur~\cite{Parrinello_Sulphur}, could be beneficial. The application of such an approach to obtain equilibrated \bo\ glass configurations at ambient conditions remains to be explored. Variations in the quench protocol, say a temperature-dependent quenching rate and/or changes in the density-temperature behavior also need to be explored to try and obtain the boroxol fraction of the glass at 300 K. Simulating the melt at 2000 K under constant pressure conditions (instead of constant volume at the experimentally reported density of 1.49 g/cc) and subsequently quenching it, under constant-NPT conditions, yields a glass at 300 K with a density of 1.5 g/cc. This value is significantly different from the density of real-world glass, which is 1.834 g/cc. The difference is not due to any inaccuracies in the MLP, but rather is due to the relatively high quenching rates used in simulations. As was presented earlier by us~\cite{debendra}, the density of the glass increases with decreasing quench rate. Given this, the deployment of 
a constant-NPT quenching procedure is not advisable, making the usage of the experimentally reported density-temperature data, a better recourse.

As per the topological constraint theory~\cite{PHILLIPS1985699, Mauro2009JCP,Smedskjaer2014,Sattler2019}, \bo\ glass is isostatic, i.e., the total number of bond and angle constraints 
equals the number of degrees of freedom, which makes it an 'ideal glass former'. The theory, however, does not take into account intermediate-range order, such as boroxol rings; thus, a network devoid of such rings and one that possesses them in abundance are equivalent structural models. The energy minimum at a 75\% boroxol fraction, as observed in this study, suggests that topological constraint theory requires modification to fully account for it; further investigation is needed.

One interesting in-silico experiment, first reported by Ferlat~\cite{Ferlat2015} and repeated in the current work, is quenching 
the melt from 2000 K at negative pressure conditions. Negative pressure reduces the density at 2000 K to values as low as 0.5 g/cc, and facilitates the growth of boroxol rings; when such a sample is quenched to room temperature (300 K), the large boroxol fraction is retained in the glassy state. These results, described in Section \ref{sec:neg-press}, affrim the close relationship between boroxol ring formation and the quenching protocol.

Furthermore, an amorphous configuration at 300 K with a density of 1.834 g/cm³ can also be obtained by quenching a low-density melt (e.g., 0.5 g/cm³) under constant-density conditions and subsequently applying pressure at room temperature to achieve a density of 1.834 g/cm³. By following this procedure, we were able to achieve a configuration with  boroxol fractions of up to 60\%. Although this method is not a physical one, it can be beneficial for the generation of amorphous \bo\ structures with high boroxol content. 

In summary, the current work has breached frontiers in the modeling of melt-quenched \bo\ glass containing large 
fraction of boroxol rings. The neural network potentials have been able to successfully capture a range of 
quantum interactions and enabled the generation of extensive MD trajectories for large system sizes. This framework critically facilitated the study of the emergence of intermediate range order in the supercooled liquid state of \bo\. Several interesting insights have been obtained on the energetics, structure, and rate estimates of the formation of boroxol rings in this network. The ML model stands poised to be deployed to undertake interesting approaches to attain an ultrastable \bo\ glass.

\section{Computational Methods}
When developing an MLP, one must be cautious regarding both the hyperparameters and the range dependence of the descriptors, ensuring that the MLP can effectively distinguish and predict the underlying structural motifs.  It is also crucial to have structures which are likely to be present in the simulations to be performed~\cite{Dellago2024JPC}. The current work employs MLPs developed within both the DeepMD-kit~\cite{deepmd,deepmdv2} as well as the MACE~\cite{mace_proc,mace_arxiv} architectures. While the former is used in most of the production MD simulations with LAMMPS~\cite{lammps}, the latter was used as a surrogate for quantum DFT, as it was quite faithful to DFT energies and forces, but was more computationally expensive than the former.

By incorporating a total of 8,652 frames into the dataset and dividing them into training and test sets with a ratio of 90:10 (train: 7778 and test: 874), the ML-31 model can effectively learn about the local environment of high boroxol frames with the desired densities. 
This dataset was used to train the ML-31 model, and the parity plot of energies of configurations and forces on atoms is presented in Figure \ref{fig:parity_plots} of the Supporting Information. Additionally, the hyperparameter testing for the model is detailed in Table \ref{tab:MLP_hyper_parm}. All details regarding the data collection for configurations with high boroxol fractions can be found in Section-\ref{sec:neg-press}.

The code to calculate the boroxol ring fraction was written in Python using networkx~\cite{networkx}. A distance cut-off of 1.7\AA\  was used to determine the B-O bonded pairs, which are, in turn, used for coordination environment and network analyses. This code only considers six-membered rings and checks the coordination numbers for boron and oxygen, to 3 and 2, respectively. The MD trajectories were analyzed using VMD~\cite{vmd} and home-grown codes.

To calculate the Single Point Energy (SPE), atom forces, and the virial of a configuration, we used the CP2K program \cite{cp2k}. The calculations employed the revised version of the Perdew-Burke-Ernzerhof (revPBE)~\cite{revPBE} exchange-correlation functional, and the valence electrons were described with a Double-Zeta Valence and Polarization MOLOPT with Short Range (DZVP-MOLOPT-SR)~\cite{molopt_basis} basis set. The core electrons and the nuclei were treated using the Goedecker-Teter-Hutter (GTH) pseudopotential \cite{GTH1, GTH2}. Grimme's D3 empirical corrections \cite{dispersion-D3}, with a cut-off of 40 \AA\ \, were applied to include dispersion contributions. 
A high electron density cut-off of 1200 Ry was used to obtain a well-converged energy. The self-consistent field (SCF) iterations were considered converged if the energy difference between successive iterations was less than $10^{-7}$ Hartree. Three-dimensional periodic boundary conditions were employed. 

\section*{Supplementary Material}
More details about the refinement of the machine learning potentials, testing with other hyperparameters, different quenching methods, and example input files are given in the supplementary material.

\section*{Data Availability}
The data that support the findings of this study are available within the article and its supplementary material. The datasets generated during the current study are available from the corresponding author upon reasonable request.

\section*{Acknowledgements}
This work was partly funded by a grant from the Board of Research in Nuclear Sciences (BRNS), India (Grant No.: 58/14/06/2020-BRNS/37060).  The support and the resources provided by the 'PARAM Yukti Facility' under the National Supercomputing Mission, Government of India, at the Jawaharlal Nehru Centre for Advanced Scientific Research, on which a part of the simulations were performed, are gratefully acknowledged. Prof. Anoop Krishnan, IIT, Delhi, is acknowledged for insightful discussions on topological constraint theory and for pointing us to Ref.~\cite{siddharth_kob_2020}.

\bibliography{ref}
\end{document}